\documentclass{iau}
\usepackage{graphicx}
\usepackage[]{natbib}  

\newcommand{\aap}{A\&A}
\newcommand{\mnras}{MNRAS}
\newcommand{\nat}{Nature}
\newcommand{\apj}{ApJ}

\title[Investigating magnetic activity of F stars with the {\it Kepler} mission] 
{Investigating magnetic activity of F stars with the {\it Kepler} mission}

\author[S. Mathur et al.]   
{S. Mathur$^1$, R.~A. Garc\'ia$^2$,  J. Ballot$^3$, T. Ceillier$^1$,  D. Salabert$^2$, T.S. Metcalfe$^1$, C. R\'egulo$^4$, A. Jim\'enez$^4$, \and S. Bloemen$^5$}

\affiliation{$^1$Space Science Institute, Boulder, CO, USA, email: {\tt smathur@spacescience.org} \\[\affilskip]
$^2$Laboratoire AIM, CEA Saclay, Gif-sur-Yvette, France
$^3$Institut de Recherche en Astrophysique et Plan\'etologie, Toulouse, France
$^4$Instituto de Astrof\'isica de Canarias, Tenerife, Spain
$^5$Radboud University Nijmegen, The Netherlands}

\pubyear{2013}
\volume{xxx}  
\pagerange{119--126}
\jname{Magnetic Fields Throughout Stellar Evolution}
\editors{A.C. Editor, B.D. Editor \& C.E. Editor, eds.}
\begin{document}

\maketitle

\begin{abstract}
The dynamo process is believed to drive the magnetic activity of stars like the Sun that have an outer convection zone.ÊLarge spectroscopic surveys  showed that there is a relation between the rotation periods and the cycle periods: the longer the rotation period is, the longer the magnetic activity cycle period will be. Ê
We present the analysis of  F stars observed by {\it Kepler}  for which individual p modes have been measure and with surface rotation periods shorter than 12 days. We defined magnetic indicators and proxies based on photometric observations to help characterise the activity levels of the stars. With the {\it Kepler} data, we investigate the existence of stars with cycles (regular or not), stars with a modulation that could be related to magnetic activity, and stars that seem to show a flat behaviour. 
\end{abstract}

\firstsection 
\section{Introduction}

Stellar magnetic activity is important to understand the solar magnetic cycle. For stars like the Sun, it results from the interaction between rotation, convection and magnetic field.  An important aspect of measuring magnetic activity in other stars is to improve the rotation-age-activity relationship, which would allow us to measure the age of a star by measuring its magnetic activity level and its rotation period in a model-independent way. The last four years, the {\it Kepler} mission has been providing very good quality data  allowing us to probe the structure \cite[e.g.][]{2012Natur.481...55B} and the dynamics \cite[e.g.][]{2012ApJ...756...19D} of stars using asteroseismology. Asteroseismology has also contributed in a more accurate determination of radius and mass of the exoplanets \citep{2012ApJ...746..123H} and the detection of magnetic activity \citep{2010Sci...329.1032G}. With long and continuous datasets provided by the mission, we now have the opportunity to study the magnetic activity of the stars.

\section{Analysis}

\noindent We used the long-cadence data (sampling of 29.42 min) corrected as described in \cite{2011MNRAS.414L...6G}.
We analysed 22 stars of spectral type F that have been observed by the {\it Kepler} mission for almost 4 years. They were selected based on their effective temperature ($T_{\rm eff} \ge$\,6000\,K) and their rotation period measurements from Garc\'ia et al. (in prep.) ($P_{\rm rot} \le$\,12 days). 
The magnetic activity measurement is based on the presence of spots or active regions on the surface of the stars. As the stars rotate, the regular passage of these dark spots produce a modulation in the light curve related to the surface rotation period of the star.  
We measured the magnetic index, $\langle S_{\rm ph}\rangle$ based on our knowledge of rotation rates of the stars. Briefly, this is the mean value of standard deviations measured on subseries of length $5 \times P_{\rm rot}$ \cite{mathur2013d}. We also performed a time-frequency analysis with the wavelets \citep{2010A&A...511A..46M} to compute a magnetic proxy. This allowed us to look for signature of magnetic activity cycles.

\section{Results}

\noindent The time-frequency analysis led to different types of magnetic activity in our sample of stars: long-lived features on the surface suggesting the existence of active longitudes,  cycle-like behaviours, some trends, and flat behaviours. We also looked for correlation between the magnetic index and other stellar parameters. We show in Fig.~1 $\langle S_{\rm ph}\rangle$ as a function of the rotation period. We do not see any correlation for the whole sample but a hint of correlation for the stars with long-lived features (triangles). But we remind that the index is taken at a given moment in the magnetic cycle of the star and might be biased depending on the observation during a maximum of minimum of magnetic activity. More details can be found in \cite{mathur2013}. The next step will be to run 3D dynamo models of a few of these stars opening the path to our understanding of the dynamics of the stars.

\begin{figure}[h!]
\begin{center}
 \includegraphics[angle=90, width=2.6in]{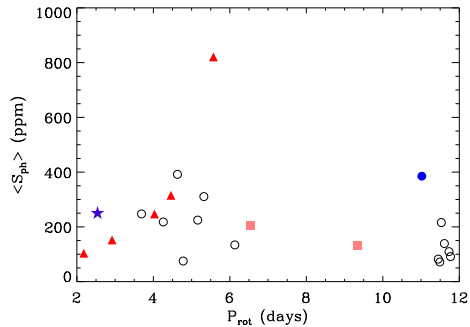} 
 \caption{Magnetic index $<S_{\rm ph}>$ as a function of  surface rotation period for the 22 stars: long-lived features (triangles), cycle-like (filled circle), trend (squares), and others (open circles). The star represents a star with long-lived features and a cycle-like behaviour. }
   \label{fig1}
\end{center}
\end{figure}

\acknowledgements{SM acknowledges support from the NASA grant NNX12AE17G. T.C., RAG, and SM acknowledge the support of the European Community's Seventh Framework Programme (FP7/2007-2013) under grant agreement no. 269194 (IRSES/ASK).}

\end{document}